\documentclass[preprint]{aastex}

\begin{document}


\title{First Observations of the
Magnetic Field Geometry in Pre-stellar Cores}


\author{D. Ward-Thompson and J. M. Kirk}
\affil{Department of Physics \& Astronomy, Cardiff University, PO Box 913,
Cardiff, UK}

\author{R. M. Crutcher}
\affil{Astronomy Department, University of Illinois, Urbana, IL 61801}
%\email{crutcher@astro.uiuc.edu}

\author{J. S. Greaves and W. S. Holland}
\affil{Joint Astronomy Center, 660 N.A`Ohoku Place, University Park,
Hilo, Hawaii 96720}

\and

\author{P. Andr\'e}
\affil{CEA, DSM, DAPNIA, Service d'Astrophysique, 
C.E. Saclay, F-91191 Gif-sur-Yvette Cedex, France}



\begin{abstract}
We present the first published maps of magnetic fields in pre-stellar cores, 
to test theoretical ideas about the way in which the
magnetic field geometry affects the star formation process.
The observations are JCMT--SCUBA maps of $\lambda 850$ $\mu$m 
thermal emission from dust. Linear polarizations at typically ten or more
independent positions in each of three objects, L1544, L183 and L43 
were measured, and the geometries of the magnetic fields in the plane 
of the sky were mapped from the polarization directions. The observed
polarizations in all three objects appear smooth and fairly uniform.
In L1544 and L183 the mean magnetic fields are at an angle of 
$\sim$30$^\circ$ to the minor axes of the cores. The L43 B-field 
appears to have been influenced in its southern half, such 
that it is parallel to the wall of a cavity 
produced by a CO outflow from a nearby T Tauri star, whilst in the
northern half the field appears less disturbed and has an angle 44$^\circ$
to the core minor axis. We briefly compare our results with published
models of magnetized cloud cores and conclude that no current model can
explain these observations simultaneously with previous ISOCAM data.
\end{abstract}




\section{Introduction}

It has become increasingly clear that magnetic fields play an important 
role in the formation and evolution of molecular 
clouds and in the process by which stars are formed.
Observations of the geometry of magnetic fields in
molecular clouds appear therefore to be essential to a full understanding
of the star formation process. Submillimetre polarimetry is one of the
most direct methods of studying magnetic field geometries in star
formation regions (e.g. Goodman 1996). 
Such observations should also allow one
to assess the relative importance of the uniform and tangled fields.
A high level of polarization, uniform in direction, indicates a
well-ordered field that is not significantly tangled on scales smaller
than the beam. Furthermore, the fact that the submm only traces the densest
material means that observations do not sample long path-lengths of magnetic
field between observer and object. One should therefore be able to
test the geometrical predictions of theoretical models.

Virtually all observations of magnetic fields in dense molecular clouds 
to date
have been toward regions of active star formation, including high-mass 
star formation sites with H~II regions. Although column densities and 
temperatures of dust grains are higher in such regions, making mapping 
of the magnetic fields more feasible, there is the danger that star
formation activity may have changed the initial geometry of the magnetic 
fields. It is therefore highly desirable to map magnetic field geometries 
in objects which are believed to be gravitationally bound, but have not yet 
formed a central hydrostatic protostar. 

Myers and co-workers (see Benson \& Myers 1989, and references therein)
identified a large number of ammonia cores, which were shown to 
be sites of low-mass star formation. \citet{BMEHMBJ86} separated the 
ammonia cores into those in which star formation had already commenced 
(those with IRAS sources) and the ``starless'' cores (those without IRAS 
sources). \citet{DWT94} observed the submillimetre continuum emission
from starless cores to ascertain their morphologies and densities and
named the most centrally condensed objects ``pre-protostellar'' (or
``pre-stellar'' for brevity) cores. \citet{AWM96} carried out a detailed 
study of the pre-stellar core L1689B to compare it to the predictions of an
ambipolar diffusion model and found some discrepancy with the
magnetic field strength required by the model. \citet{DWT99}
observed the millimetre continuum emission from a larger
number of pre-stellar cores
to ascertain their detailed radial density profiles, and confirmed
the central flattening that had previously been observed by \citet{DWT94}.
A small number of cores
are sufficiently bright at $\lambda 850$ $\mu$m for polarization 
mapping observations to be currently
possible. In this paper we present the first published submm continuum
polarization observations of pre-stellar cores, which were carried out 
to test the predictions of theoretical models.

\section{Observations}

Submillimetre continuum polarimetry observations at $\lambda$850~$\mu$m were
carried out using the SCUBA (Submillimetre Common User Bolometer Array)
camera on the James Clerk Maxwell Telescope (JCMT)\footnote{JCMT 
is operated by the Joint Astronomy Center, Hawaii, on 
behalf of the UK PPARC, the Netherlands NWO, and the Canadian NRC.
SCUBA was built at the Royal Observatory, Edinburgh. SCUBAPOL was built at
Queen Mary \& Westfield College, London.} 
on Mauna Kea, Hawaii, on the mornings of 1999
March 15 (L183 \& L43) and September 15 (L1544) from HST 01:30 to 09:30
(UT 11:30 to 19:30).
SCUBA was used in conjunction with the polarimeter SCUBAPOL,
in the 16-position
jiggling mode to produce a fully sampled 2.3 arcmin image
(Holland et al. 1998).

The observations were carried out while using the secondary mirror 
to chop 120
arcsec in azimuth at around 7 Hz and synchronously to detect the signal, 
thus rejecting `sky' emission. 
The method of observing used was to make a full
16-point jiggle map (to produce a Nyquist sampled map), with
an integration time of 1 second per point, 
in each of the `left' and `right' beams of the telescope
(the two beams are produced by the chopping secondary).
This process is repeated at each position of the
polarimeter half-wave plate. Then the half-wave plate
is rotated
to the next position, in steps of 22.5$^\circ$. 16 such positions
thus constitute a complete revolution of the half-wave plate,
representing 512 seconds of on-source integration, which takes about
12 minutes, including overheads
(see Greaves 1999, and
Greaves et al. 2000, for a description of SCUBAPOL).
This process was then repeated
several times -- 14 for L183, 15 for L43 and 9 for L1544.
The instrumental polarization (IP) of each bolometer
was measured on the planets Uranus
and Saturn, and subtracted from the data before calculating the true
source polarization. The mean IP was found to be 0.93$\pm$0.27\%.
The observations were repeated with a slight offset in each case
so that the three bolometers 
with significantly above average noise could be 
removed without leaving areas of the map with no data.
The atmospheric opacity at 225~GHz was monitored
by the radiometer located at the Caltech Submillimeter Observatory.
The opacity at 225~GHz was 0.06 during the L43 and L183 observations, 
and 0.07 during the L1544 observations, typical of fairly good 
conditions at the site, and corresponding  to a zenith atmospheric 
transmission at $\lambda$850~$\mu$m of about 80\%. 

Subsets of the data were reduced separately and it was seen that 
atmospheric stability, rather than purely transparency, was the limiting 
factor to the repeatability of the results obtained. Those data that were
taken during unstable periods were seen to generate only noise, whereas
all data taken during stable periods produced consistent results and were
co-added to produce the maps shown here.
During earlier observing runs with SCUBAPOL, when it was operating only in 
single-pixel mode, on 1998 February 13th \& 14th, and March 2nd \& 4th (UT),
we observed the central peak positions of the same three sources.
We found the current results agreed to within the errors
with these previous measurements, confirming the repeatability of our results.
Pointing and focussing were checked using the bright sources Uranus 
and IRAS16293 for L183 and L43, and CRL618 for L1544, and the pointing 
was found to be good to $\sim$1--2~arcsec throughout.

\section{Results}

Figures 1-3 present the results of our observations. In each case the
Stokes I map of dust emission is shown as a grey-scale with
contours overlaid.
The direction of {\bf B} in the plane of the sky is shown by a series of
vectors at every position where a measurement of the polarized flux above 
the 2-$\sigma$ level was achieved. Most vectors have a
much higher signal-to-noise ratio than this, up to a maximum of 12$\sigma$,
with a mean of $\sim$5$\sigma$ across all three sources. 
The 2-$\sigma$ lower cutoff implies a maximum uncertainty 
in any given position angle of $\pm$14$^\circ$. Note that we refer 
to these as magnetic field vectors, but they are not true vectors 
because they have a 180$^\circ$ directional ambiguity. 
The plotted B-vectors 
are perpendicular to the direction of the polarization observed, in accord 
with all of the various paramagnetic relaxation mechanisms of grain 
alignment by interstellar magnetic fields (e.g. Purcell 1979), 
and the length of each 
B-vector is proportional to the percentage polarization.
Hence we are making the assumption that the polarization we are mapping
is tracing the magnetic field direction. Given this
reasonable assumption, we deduce the field direction in each source.
The scale is such that a vector of length 2~arcsec represents a
percentage polarization of 1\%. The vectors are 
plotted on a grid spacing equal to the diameter of the JCMT beam, so each 
vector is independent.

\subsection{L1544}

Figure 1 shows our observations of the L1544 core,
where we follow the naming convention of \citet{BM89} and 
refer to the only known
pre-stellar core in the larger dark cloud L1544 as the L1544 core.
Our $\lambda 850$ $\mu$m data give core full width at half maximum 
(FWHM) dimensions of 
$\sim$110$\times$60 arcsec, with the minor axis
at a position angle of 52$\pm$5$^\circ$ (all angles are measured
north through east). The core is at a distance of 140~pc, and
we have previously measured its mass to be
3.2$M_\sun$ (Ward-Thompson et al. 1999). \citet{BM89}
found that $\Delta V(NH_3) = 0.3$ km s$^{-1}$, 
which implies a virial mass of $\sim 1.7$ $M_\sun$. Hence this core
appears to be gravitationally bound. Furthermore, recent spectroscopic
observations show that the core appears to be contracting, although
it is not apparently undergoing free-fall collapse \citep{TAFF}.
The weighted mean position angle of the magnetic 
field, averaged over the 8 positions, is 23$\pm$2$^\circ$.
Hence we see that the mean magnetic field direction is not along the 
minor axis of the pre-stellar core, but
at an angle of 29$\pm$6$^\circ$ to the minor axis,
and the field appears to be fairly uniform.
There is also some evidence that the percentage polarization decreases
towards the peak of the source -- this is discussed further in section 4
below.

There has been some recent debate about the nature of the physical
processes which are currently determining the evolution of L1544.
Ward-Thompson et al. (1999) showed that L1544 has a radial density 
profile which is
consistent with that predicted by ambipolar diffusion theory, whilst
noting some inconsistencies in the apparent time-scales. \citet{WMWD99}
claimed a discrepancy between the observed infall motions and the
predictions of ambipolar diffusion theory. Subsequently, \citet{CB99}
have shown that in fact the ambipolar diffusion model can be fine-tuned
to explain these motions. However, this model still requires that the
magnetic field lies parallel to the minor axis of the core and
our observations disagree with this. 

\subsection{L183}

Figure 2 shows our observations of the L183 core, where
we once again refer to the only
known starless core in the L183 dark cloud simply as the L183 core
(it should be noted, however, that the L183 cloud is sometimes also
referred to as L134N). Our $\lambda 850$ $\mu$m 
data give core FWHM dimensions of $\sim$120$\times$60 arcsec (although
this is not so clearly defined, since the overall
source extent north-south is greater than the SCUBA field of view),
with the minor axis at a position angle of 
80$\pm$5$^\circ$. The core is at a distance of 150~pc, and
we measured its mass to be $\sim$1.3~$M_\sun$ (Ward-Thompson et al. 1999). 
It has previously been seen that this core has
$\Delta V(NH_3) = 0.24$ km s$^{-1}$ \citep{BM89}, 
implying a virial 
mass of $\sim$1.2$M_\sun$. Hence this core 
also appears to be gravitationally
bound.

The weighted mean position angle of the magnetic field,
averaged over the 16 positions, is 46$\pm$2$^\circ$.
Thus, the mean magnetic field is not parallel to
the minor axis of the pre-stellar 
core, but at an angle 34$\pm$6$^\circ$ to it, very similar to
the situation observed in L1544. Once again,
the magnetic field direction is fairly constant, and
all variations are consistent with measurement uncertainties.
Here also there is some evidence for the percentage polarization to
decrease at the highest intensities (see section 4).

\subsection{L43}

Figure 3 illustrates our
observations of the starless core in the L43 dark cloud.
The $\lambda 850$ $\mu$m data give the core FWHM dimensions
to be $\sim$100$\times$60 arcsec,
with its minor axis at position angle 37$\pm$5$^\circ$.
It lies at a distance of 170~pc, and our separate SCUBA
observations imply that 
its mass is $\sim$4~$M_\sun$. \citet{BM89} found that
$\Delta V = 0.4$ km s$^{-1}$, which 
gives a virial mass of $\sim$2.1$M_\sun$, once again apparently
showing that it is gravitationally bound.
However, this is a more complicated 
region than the others. Figure 3 shows a second core at the 
western edge of the SCUBA field of view. There is a 
classical T Tauri star, RNO~91, embedded in this second core \citep{COH80}, 
and a molecular outflow centred on this source \citep{MBFMS88}.

Study of the pre-stellar core in the centre of the field shows a general area
of roughly uniform polarization at position angle
roughly 170$^\circ$ extending from
the centre of the core to the north. However, to the south and west of the
core centre there is an area of vectors with position angle
roughly 140$^\circ$.
Moving still further to the west, the B-field appears to turn smoothly
through an angle of roughly 90$^\circ$ until it has a position angle of
233$\pm$4$^\circ$ 
(=53$\pm$4$^\circ$) at the position of RNO~91 at the western edge of 
the field.

We interpret this as follows: The outflow from RNO~91 is known to have
cleared a cavity to the south of the source and the southern edge of the
pre-stellar core forms part of the cavity wall (c.f. Bence et al. 1998).
Hence we believe that the B-field we observe in the southern and western
parts of the pre-stellar core has in all probability been affected by the
interaction with the outflow, and lies roughly parallel to the cavity
wall. However, away from the influence of the 
outflow, in the northern part of the pre-stellar core, the B-field we
observe is more likely to represent the initial unperturbed field
direction in the pre-stellar core. If this is the case, then 
the weighted mean B-field, averaged over these 9 vectors,
has a position angle 173$\pm$2$^\circ$, and the angle between the B-field and
the core minor axis is 44$\pm$6$^\circ$. This is similar to
the angle offset we observe in each of the other two cores.
However, some caution must be attached to this estimate, since at the
core centre itself the percentage polarization is very low.
In fact the L43 pre-stellar core shows the clearest evidence for 
depolarization towards the core peak (see next section for discussion).

We also note that the magnetic field we observe in the RNO~91 region is 
perpendicular to the elongation axis of the edge-on
circumstellar disk around RNO~91 that was
discovered by the optical polarimetry observations of \citet{SDT93}.
However, if this
disk originally collimated the bipolar outflow from this source, 
then the outflow must have subsequently turned through almost 90$^\circ$
in breaking out of the cloud, as it is now seen to extend
to large distances in an almost north-south direction 
\citep{BPIWW98}.

\section{Discussion and Conclusions}

Theoretical models make predictions about the strength and geometry 
of magnetic fields that can be tested by observations. 
A crucial parameter in any theory or simulation of the structure and 
evolution of magnetic clouds is the ratio of thermal to magnetic
pressures, $\beta$ (c.f. Ostriker et al. 1999; Crutcher 1999).
For example, \citet{OGS99} follow the evolution of an initially 
uniform medium for various values of
$\beta$ subjected to perturbations. For 
$\beta > 1$ turbulence dominates and the field lines become heavily 
tangled.

In the magnetically dominated case ($\beta << 1$), clouds form 
by material streaming along field lines, so that structures that are 
elongated preferentially perpendicular to the magnetic field are 
formed and the field geometry is uniform, without 
small-scale random variations. Most such models of 
the evolution of self-gravitating molecular cores supported by 
magnetic fields (e.g. Ciolek \& Mouschovias 1994;
Li \& Shu 1996) predict that the cores should be oblate 
spheroids with minor axes parallel to the magnetic field direction. 

The cores then evolve by the process of ambipolar
diffusion, in which the neutral gas diffuses through the ionised
component, which is held static by the magnetic field. However, 
simulations show that the magnetic field need not always be 
perpendicular to the elongation. If $\beta = 0.1\rightarrow 1$, then
significant deviations are often found
\citep{OGS99}, although
the initially uniform magnetic field remains the dominant component.

We have mapped the $\lambda$850~$\mu$m polarization in three pre-stellar
cores, L1544, L183 \& L43, and used these measurements to infer the
magnetic field geometries of these objects. In L1544 and L183, as well as
in the region of L43 that we believe to be undisturbed by the nearby outflow,
we see relatively uniform polarizations, leading us to infer
a uniform magnetic field direction in each case.
By comparison with the above-mentioned models, this suggests
that those models which assume $\beta\leq 1$ are
applicable to these regions (e.g. Ciolek \& Mouschovias 1994; Li \& Shu 1996;
Ciolek \& Basu 2000). Our observation in each case of an offset 
between the position angle of the B-field and the minor axis of each core
appears to suggest that $\beta > 0.1$ (c.f. Ostriker et al. 1999). 

Some level of depolarization is observed towards the peak of each
pre-stellar core. A similar depolarization effect was seen
in OMC-3 by Matthews \& Wilson (2000).
This may be indicating that the field has
small scale structure below our resolution limit, causing the observed
percentage polarization to decrease. If this is the case, then we can
estimate the amount by which the random component diverges from the (larger)
uniform component. Following Hildebrand \& Dragovan (1995) we estimate
that if the amount of depolarization is up to a factor of $\sim$2, as we
observe, then the small scale random field could diverge from the ordered
field by up to $\sim$35$^\circ$. However, given the relatively
high levels of polarization that we still observe at the core peaks,
we believe this is an upper limit. The 
alternative explanation for depolarization is that
in denser regions the number of collisions increases and hence 
the grain alignment efficiency decreases. 
We favour the latter explanation because we do not
see the polarization direction varying, suggesting that the uniform ordered
B-field component is still dominant, in agreement with our above deduction
that $\beta < 1$.

Recent ISOCAM observations of pre-stellar cores seen in absorption 
at 7 \& 15~$\mu$m, have found that a number of cores, including L1544, 
have very sharply defined edges (Bacmann et al. 2000). From comparison with 
various ambipolar diffusion models, Bacmann et al. concluded that only 
cloud cores which are highly magnetically subcritical initially can 
develop such sharp edges. For example, the best fit model to the ISOCAM data 
of L1544 has $\beta\sim$0.08 initially (c.f. Ciolek \& Mouschovias 1995), in 
apparent contradiction with the findings of this paper. On the other hand, 
the model of L1544 proposed by Ciolek \& Basu (2000) which has an initial 
$\beta\sim$0.2, in better agreement with our present conclusions, 
cannot explain the ISOCAM data. Similarly the models of Ostriker et
al. (1999) cannot simultaneously explain the polarization data in this paper 
(requiring $\beta > 0.1 $) and the large density contrasts associated 
with pre-stellar cores (suggesting $\beta < 0.1 $). 

Hence, our comparison of the observations and theoretical results leads us
to conclude that no current model of magnetically regulated star formation 
can apparently account for all of the existing observations. 

\acknowledgments

J.M.K. wishes to acknowledge PPARC for studentship support.
R.M.C. is partially supported by National Science Foundation grant 
AST9820641 to the University of Illinois.
We would also like to thank the JCMT telescope operators for their
assistance during these observations, and Tim Jenness for assistance
with the software.




%% Generally speaking, only the figure captions, and not the figures
%% themselves, are included in electronic manuscript submissions.
%% Use \figcaption to format your figure captions. They should begin on a
%% new page.

\clearpage

%% No more than seven \figcaption commands are allowed per page,
%% so if you have more than seven captions, insert a \clearpage

%% There must be a \figcaption command for each legend. Key the text of the
%% legend and the optional \label in curly braces. If you wish, you may
%% include the name of the corresponding figure file in square brackets.
%% The label is for identification purposes only. It will not insert the
%% figures themselves into the document.
%% If you want to include your art in the paper, use \plotone.
%% Refer to the on-line documentation for details.

\figcaption[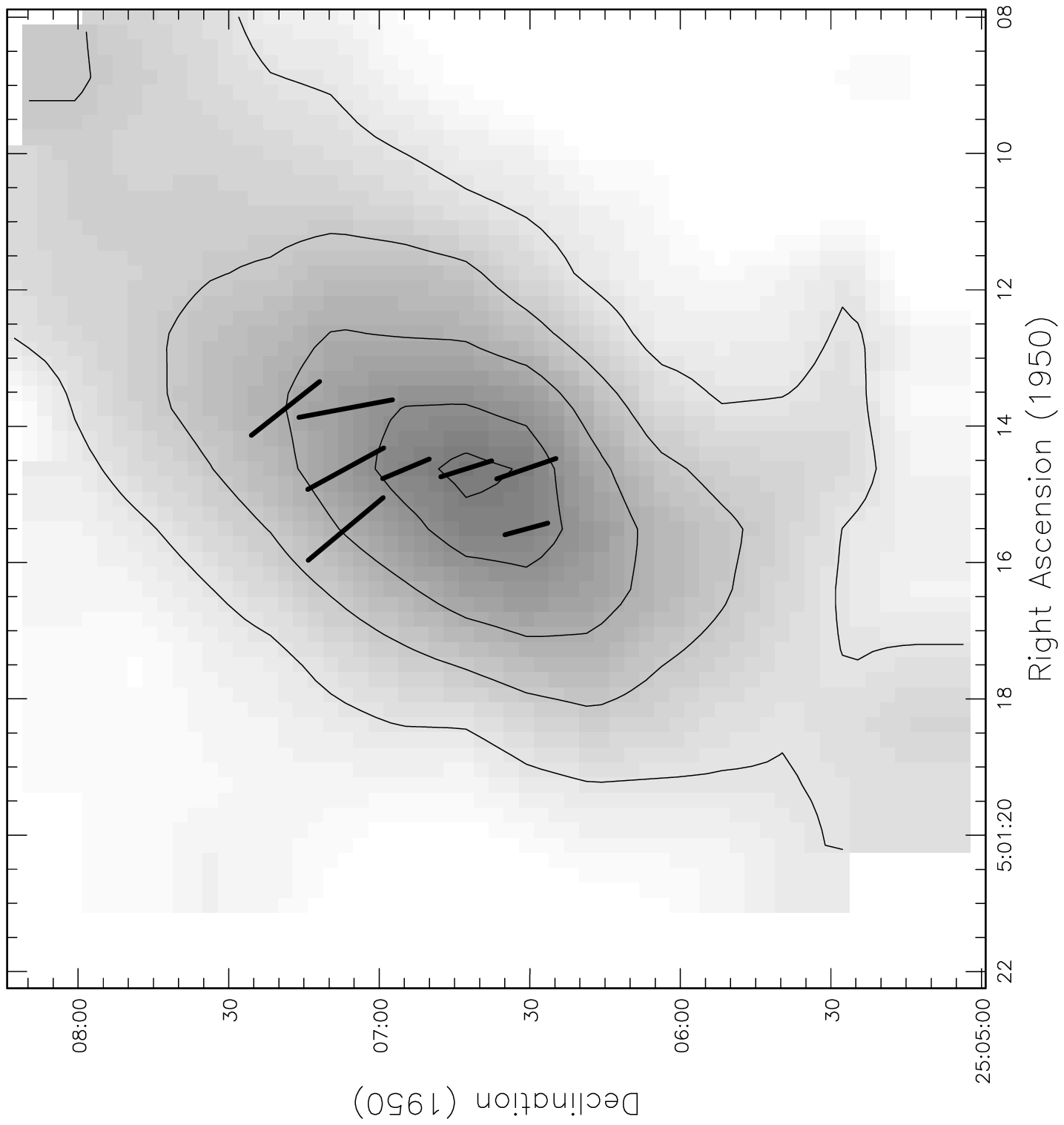]{Dust continuum emission at $\lambda 850$ $\mu$m 
from the L1544 pre-stellar core. The
Stokes I map is shown as a grey-scale with
contours overlaid. Contour levels are at 20, 40, 60, 80 \& 95\% of peak.
The direction of the B-field in the plane of the sky is shown by a series 
of vectors at every position where a measurement of the polarized flux above 
the 2-$\sigma$ level was achieved. The plotted B-vectors 
are perpendicular to the direction of the polarization observed,
and the length of each 
B-vector is proportional to the percentage polarization,
such that a vector of length 2~arcsec represents a
percentage polarization of 1\%. The vectors are 
plotted on a grid spacing equal to the diameter of the JCMT beam, so each 
vector is independent. 
\label{fig1}}

\figcaption[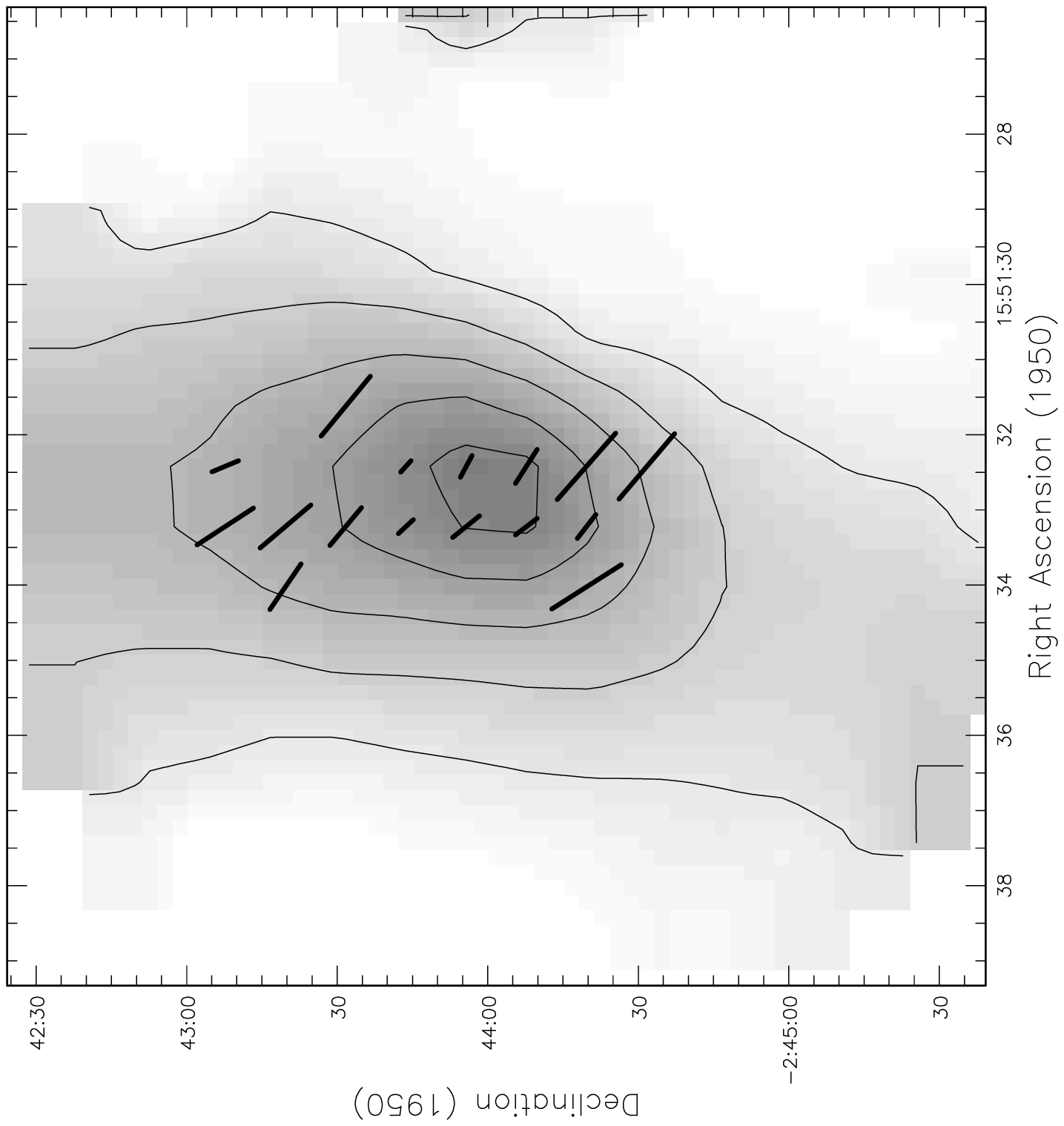]{Dust continuum emission at $\lambda 850$ $\mu$m 
from the L183 pre-stellar core, with vectors of the inferred B-field
direction overlaid. Details as in Figure 1.
\label{fig2}}

\figcaption[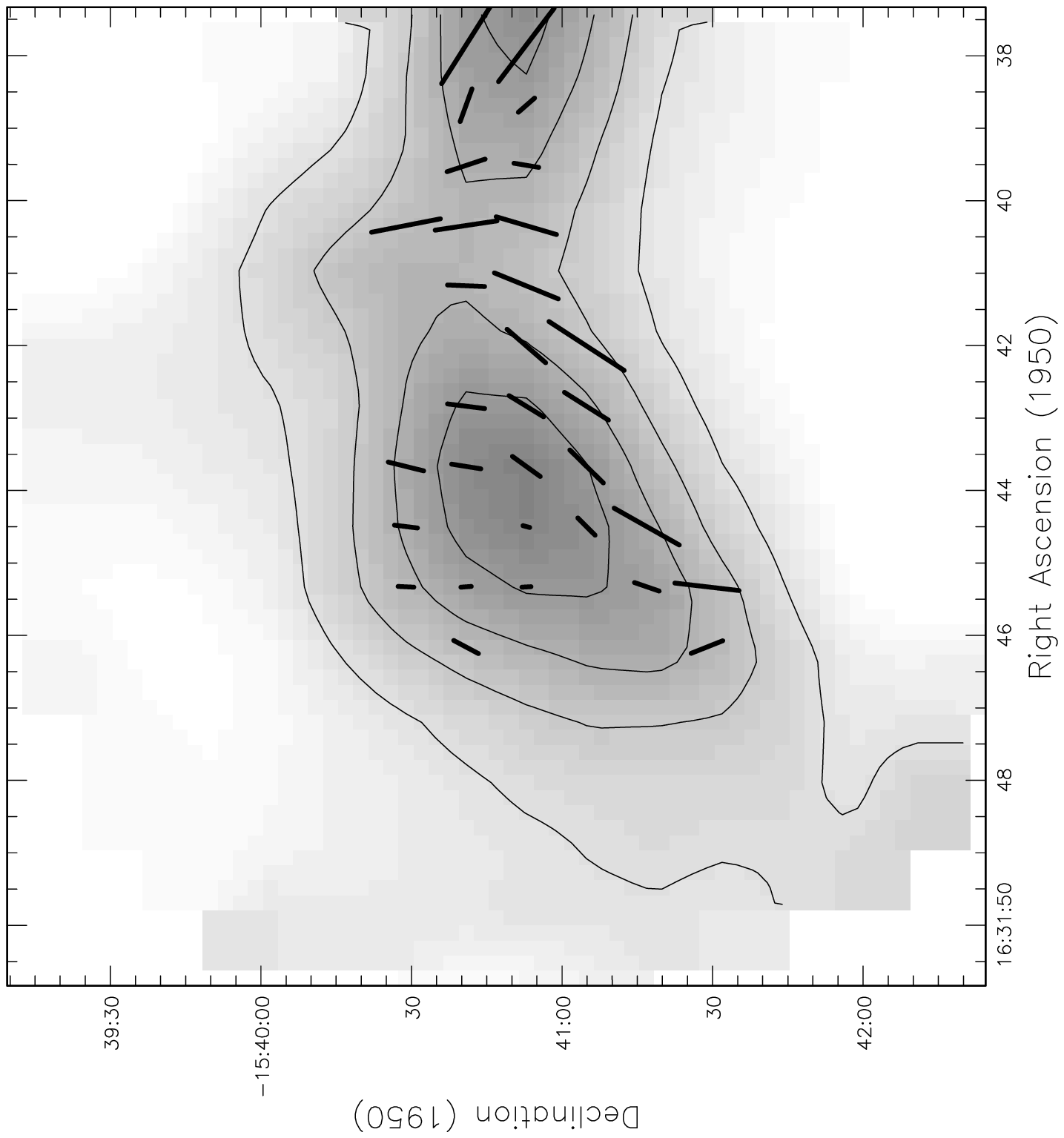]{Dust continuum emission at $\lambda 850$ $\mu$m from 
the L43 pre-stellar core, with vectors of the inferred B-field
direction overlaid. Contour levels are at 20, 40, 60 \& 80\% of peak.
Other details as in Figure 1.
\label{fig3}}

\begin{figure}
\setlength{\unitlength}{1mm}
\begin{picture}(80,120)
\includegraphics{fig1.ps}
\end{picture}
\end{figure}

\begin{figure}
\setlength{\unitlength}{1mm}
\begin{picture}(80,120)
\includegraphics{fig2.ps}
\end{picture}
\end{figure}

\begin{figure}
\setlength{\unitlength}{1mm}
\begin{picture}(80,120)
\includegraphics{fig3.ps}
\end{picture}
\end{figure}


\begin{thebibliography}{99}

\bibitem[Andr\'e, Ward-Thompson \& Motte (1996)]{AWM96}
Andr\'e P., Ward-Thompson D., Motte F., 1996, A\&A, 314, 625

\bibitem[Bacman et al. (2000)]{BAC00} Bacmann A., Andr\'e P., Puget J-L., 
Abergel A., Bontemps S., Ward-Thompson D., 2000, A\&A, in press

\bibitem[Beichman et al. (1986)]{BMEHMBJ86} Beichman C. A., Myers P. C., 
Emerson J. P., Harris S., Mathieu R., Benson P. J., Jennings R. E. 
1986, \apj, 307, 337

\bibitem[Bence et al. (1998)]{BPIWW98} Bence, S. J., Padman, R., Isaak, 
K. G., Wiedner, M. C., Wright, G. S. 1988, \mnras, 299, 965

\bibitem[Benson \& Myers (1989)]{BM89} Benson, P.J., Myers, P.C 1989, 
\apjs, 71, 89

\bibitem[Ciolek \& Basu (2000)]{CB99} Ciolek G. E., Basu S.,
2000, \apj, 529, 925

\bibitem[Ciolek \& Mouschovias (1994)]{CM94} Ciolek G. E., Mouschovias 
T. Ch., 1994, \apj, 425, 142

\bibitem[Ciolek \& Mouschovias (1995)]{CM95} Ciolek G. E., Mouschovias 
T. Ch., 1995, \apj, 454, 194

\bibitem[Cohen (1980)]{COH80} Cohen M., 1980, AJ, 85, 29

\bibitem[Crutcher (1999)]{Crutch} Crutcher R. M., 1999, ApJ, 520, 706

\bibitem[Greaves et al. (1999)]{G99} Greaves, J., 1999, 
``The SCUBA polarimeter users' guide'', version 4, Joint Astronomy
Center, Hawaii

\bibitem[Goodman (1996)]{Go96} Goodman A. A., 1996, in: `Polarimetry of 
the interstellar medium', eds: Roberge W. G., Whittet D. C. B., ASP
Conference Series, 97, 325

\bibitem[Greaves et al. (2000)]{G20} Greaves J. S., Jenness T., 
Chrysostomou A. C., Holland W. S., Berry D. S., 2000, in ``Imaging at 
Radio through Submillimeter Wavelengths'', ed: Mangum J.,
ASP Conference Series, in press

\bibitem[Hildebrand \& Dragovan (1995)]{HD95}
Hildebrand R. H., Dragovan M., 1995, ApJ, 450, 663

\bibitem[Holland et al. (1998)]{H98} Holland W. S., Robson E. I., 
Gear W. K., Cunningham C. R., Lightfoot J. F., Jenness T., Ivison R. J., 
Stevens J. A., Ade P. A. R., Griffin M. J., Duncan W. D., Murphy J. A., 
Naylor D.A., 1999, MNRAS, 303, 659 

\bibitem[Li \& Shu (1996)]{LS96} Li Z-Y., 
Shu F. H., 1996, \apj, 472, 211

\bibitem[Mathieu et al. (1988)]{MBFMS88} Mathieu R. D., Benson P. J., 
Fuller G. A., Myers P. C., Schild R. E. 1988, \apj, 330, 385

\bibitem[Matthews \& Wilson (2000)]{MW20}
Matthews B. C., Wilson C. D., 2000, ApJ, 531, 868

\bibitem[Ostriker et al. (1999)]{OGS99} Ostriker E. C., Gammie C. F., 
Stone J. M. 1999, \apj, 513, 259

\bibitem[Purcell (1979)]{Purc} Purcell E. M., 1979, ApJ, 231, 404

\bibitem[Scarrott, Draper, \& Tadhunter (1993)]{SDT93} Scarrott, S. M., 
Draper, P. W., Tadhunter, C. N. 1993, \mnras, 262, 306

\bibitem[Tafalla et al. (1998)]{TAFF} Tafalla M., Mardones D., Myers P. C.,
Caselli P., Bachiller R., Benson P. J., 1998, \apj, 504, 900

\bibitem[Ward-Thompson et al. (1994)]{DWT94} Ward-Thompson, D., Scott, P.F., 
Hills, R.E., Andr\'e, P. 1994, \mnras, 268, 276

\bibitem[Ward-Thompson, Motte, \& Andr\'e (1999)]{DWT99} Ward-Thompson, D., 
Motte, F., Andr\'e, P. 1999, \mnras, 305, 143

\bibitem[Williams et al. (1999)]{WMWD99} Williams J. P., Myers P. C., 
Wilner D. J., Di Francesco J., 1999, \apj, 513, L61

\end{thebibliography}
\end{document}